\documentclass[11pt,twoside]{article}
\usepackage{asp2004}
\usepackage{psfig}
\usepackage{epsf}
\usepackage{graphics}
\usepackage{lscape}
\def\edcomment#1{\iffalse\marginpar{\raggedright\sl#1\/}\else\relax\fi}
\reversemarginpar

\begin{document}
\title{Chemical Abundances  and Yields from Massive Stars }
\author{Andr\'e  Maeder, Georges Meynet \& Raphael Hirschi}
\affil{Geneva Observatory, CH-1290 Sauverny, Switzerland}

\begin{abstract}
Stellar rotation produces an internal mixing of the elements due to shear instability and meridional
circulation. This  leads to observable $N/C$
 enhancements in massive stars above about 7--9 $M_{\odot}$. Rotation also favours mass loss
 by stellar winds. Mass loss effects dominate for masses above 30 $M_{\odot}$, while
 mixing  dominates below that limit.
 The effects of mixing are also  much larger at lower metallicity $Z$, because the internal 
 $\Omega$--gradients are steeper. This appears to be in agreement with observations 
 in the SMC.
 
 At very low $Z$ and $Z=0$, mixing between the He--burning core and the H--burning shell
 leads to the production of primary N in intermediate mass stars.  
 Such enrichments increase the metallicity of the rotating star,
 also massive $Z=0$ stars with  moderate 
 initial velocities  currently reach break--up velocity during a fraction 
 of the MS phase. Both effects favour mass loss in $Z=0$ stars, which  have ejecta with 
 abundance anomalies very similar to those of C--rich very metal poor stars.

\end{abstract}
\thispagestyle{plain}

\section{Introduction}

David Lambert has made major contributions to our knowledege of chemical
abundances in all kinds of stars.  Chemical abundances are the most decisive
test on mixing in stellar evolution,  which affects all model outputs: 
tracks, lifetimes, stellar winds, supernovae 
progenitors, star populations in galaxies, nucleosynthesis and  yields, etc. Thus, the
abundance determinations made by David Lambert are an essential  piece of our astrophysical knowledge.

Here, we examine the case of massive stars, where rotation and mass loss
play a major role.  
 Noticeably, several results 
of models at low metallicity $Z$ also have 
implication for the abundances of very metal poor  halo stars.

\section{Basic physical ingredients concerning rotation and mass loss}

We briefly summarize  the phyical assumptions about the 
treatment of mass loss and rotation in stellar models. 

{\emph{Structure:}} We consider
the case of shellular rotation with $\Omega$ being  a function $\Omega(r)$ of the radius only
  \citep{Z92}, because in a differentially rotating star the horizontal turbulence is
 strong enough to ensure a constancy of $\Omega$ in latitude.  For  differentially rotating
 stars, the 
structure equations need  to be  properly written \citep{MMI}, which is often not
the case in current literature. 
Structural effects due to the centrifugal force
are in general small in the interior, while the  distorsion of stellar surface may be large
enough to produce significant shifts in the HR diagram  \citep{MP70}.

{\emph{Internal transports of chemical elements and angular momentum:}}
For these two  transports, we consider the effects of
shear mixing, of meridional circulation and their interactions  
with  horizontal turbulence. The equation of  evolution 
of chemical abundances due to the transport and nuclear processes 
is at each level in lagrangian coordinates 
\citep{Z92},

\begin{eqnarray}
 \left( \frac{dX_i}{dt} \right)_{M_r} = 
& \left(\frac{\partial  }{\partial M_r} \right)_t
\left[ (4\pi r^2 \rho)^2 (D_{\rm shear}+ D_{\rm eff}) \left( \frac{\partial X_i}
{\partial M_r}\right)_t
\right] + \left( \frac{dX_i}{dt} \right)_{\mathrm{nucl}}.
\end{eqnarray} 

\noindent
The equations of transport of angular momentum in the vertical direction is
\begin{eqnarray}
\rho \frac{d}{d t}
\left( r^2 \Omega\right)_{M_r} = 
 \frac{1}{5 r^2}  \frac{\partial}{\partial r}
\left(\rho r^4 \Omega \; U(r) \right)
  + \frac{1}{r^2} \frac{\partial}{\partial r}
\left(\rho D_{\mathrm{shear}}\; r^4 \frac{\partial \Omega}{\partial r} \right) .
\end{eqnarray} 

\noindent $\Omega(r)$ is the mean angular velocity at level $r$. $U(r)$ is the
 radial term of the vertical component of the velocity of the meridional
circulation. $D_{\mathrm{shear}}$ is the coefficient of shear diffusion. From the two terms in the 
second member in this last equation, we see that {\emph{advection and diffusion are not the same}. 
Contrarily to what is often made in recent literature,  
the transport of angular momentum by circulation cannot be 
treated as a diffusion, (by doing so one may even have the wrong sign for the
transport of angular momentum !).
For the changes of the chemical 
elements due to transport, we see that the  diffusion coefficient 
 is  the sum  of  $D_{\mathrm{shear}}$ 
and   $D_{\rm eff} = \frac{\mid rU(r) \mid^2}{30 D_h}$. $D_{\rm eff}$ expresses the resulting
effect of meridional circulation and of a large horizontal turbulence
 \citep{ChabZ92}. This expression of $D_{\rm eff}$ tells us that the vertical
advection of chemical elements is  inhibited by the
strong horizontal turbulence characterized by $D_{\rm{h}}$. 
 The usual estimate of the coefficient $D_{\mathrm{h}}$ of horizontal turbulence
 was  $D_{\mathrm{h}} =
\frac{1}{c_{\mathrm{h}}} r \;|2V(r) - \alpha U(r)| $  \citep{Z92}. 
More recent estimates give higher values for this coefficient \citep{MDh}, confirmed by further
studies \citep{Mathis04}.
 The diffusion coefficient $D_{\mathrm{shear}}$ by shear mixing essentially behaves like
 
 \begin{eqnarray}
D_{\mathrm{shear}} =  \frac{ (K + D_{\mathrm{h}})}
{\left[\frac{\varphi}{\delta} 
\nabla_{\mu}(1+\frac{K}{D_{\mathrm{h}}})+ (\nabla_{\mathrm{ad}}
-\nabla_{\mathrm{rad}}) \right] }
 \frac{H_{\mathrm{p}}}{g \delta}  
\left [ \alpha\left( p \Omega{d\ln \Omega \over d\ln r} \right)^2
-4 (\nabla^{\prime}  -\nabla) \right] 
\end{eqnarray}

\noindent 
with p=0.884. $D_{\mathrm{shear}}$
is also modified by the horizontal turbulence  $D_{\mathrm{h}}$, in the sense that a larger
$D_{\mathrm{h}}$ leads to a  decrease of the mixing. The radial component $U(r)$  of the meridional circulation is \citep{MZ98},

\begin{eqnarray}
U(r) &=&  \frac{P}{\overline{\rho} \overline{g} C_{\!P} \overline{T}
\, (\nabla_{\rm ad} - \nabla +  (\varphi/\delta) \nabla_{\mu}) }
\times   {  \frac{L}{M_\star} \left[E_\Omega + E_\mu \right] + \!
\frac{C_P}{\delta} \frac{\partial \Theta }{\partial t} \! } ,
\end{eqnarray}

\begin{figure}[!ht]
\plotfiddle{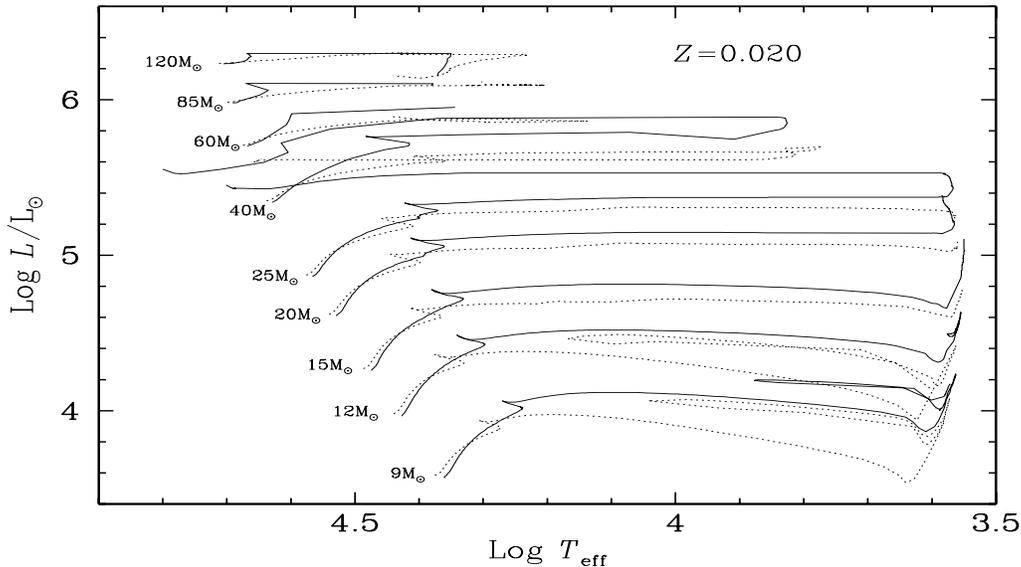}{7.0cm}{270}{50}{40}{-200}{230}
\caption{HR diagram of massive stars with Z=0.02, for rotating stars with
$v_{\mathrm{ini}}=300$ km/s (continuous lines) and for non--rotating stars (dotted lines). }
\end{figure}
\noindent
where $M_{\star}=M \left( 1 - \frac{\Omega^2}{2 \pi G \rho_{\rm{m}}}  \right)$
is the reduced mass,
with the  notations  given in the quoted paper.
The driving term in the square brackets in the second member
 is $E_{\Omega}$.  It behaves mainly like
$E_{\Omega}  \simeq  \frac{8}{3} \left[ 1 - \frac{{\Omega^2}}
{2\pi G\overline{\rho}}\right] \left( \frac{\Omega^2r^3}{GM}\right)$
The term $\overline{\rho}$ means the average
on the considered equipotential.
The term with the minus sign in the square bracket is the 
Gratton--\"{O}pik term, which becomes important in the outer layers
when the local density is small. This term  produces negative values of $U(r)$, meaning that 
the circulation is going down along the polar axis and up 
in the equatorial plane. This makes an outward transport 
of angular momentum, while a positive $U(r)$ gives an inward transport.
At lower $Z$, the Gratton--\"{O}pik term is negligible, which contributes
to make larger $\Omega$--gradients in lower $Z$ stars. 

{\emph{Mass loss in rotating stars:}} 
Rotation certainly influences the mass loss rates by stellar winds
\citep{HL98}.
For a rotating star, one must consider 
the flux $F(\vartheta)$ at a given colatitude $\vartheta$ as given by von Zeipel's theorem.
 Thus, in a rotating star, the Eddington factor becomes a local quantity $\Gamma_{\Omega}(\vartheta)$.
 We define it  as
the ratio of the local flux $F(\vartheta)$ 
given by the von Zeipel theorem to the maximum possible local flux,
which is $F_{\mathrm{lim}}(\vartheta) = - \frac{c}{\kappa(\vartheta)}  
g_{\mathrm{eff}}(\vartheta)$. Thus, one has \citep{MMVI}
\begin{eqnarray}
\Gamma_{\Omega}(\vartheta) =
\frac{F(\vartheta)}{F_{\mathrm{lim}}(\vartheta)}=
\frac{ \kappa (\vartheta) \; L(P)[1+\zeta (\vartheta)]}{4 \pi 
cGM \left( 1 - \frac{\Omega^2}{2 \pi G \rho_{\rm{m}}}  \right) } \; ,
\end{eqnarray}
\noindent 
where the opacity $\kappa(\vartheta)$ depends on the colatitude $\vartheta$, 
since $T_{\mathrm{eff}}$
also depends on $\vartheta$. 
This shows that the maximum luminosity of a rotating star is reduced by rotation.
It is to be stressed that if the limit  $\Gamma_{\Omega}(\vartheta) = 1 $ 
happens to be met in general at the equator, it is not because 
$g_{\mathrm{eff}}$ is the lowest there, but because the 
opacity is the highest!

Often, the critical velocity in a rotating star is written as
$v^2_{\rm{crit}} = \frac{GM}{R} (1-\Gamma)$. This expression is incorrect,
since it would apply only to uniformly bright stars. 
The critical velocity of a rotating star is given by the zero of the equation
expressing the total gravity 
$\vec{g_{\mathrm{tot}}} = \vec{g_{\mathrm{eff}}} + \vec{g_{\mathrm{rad}}} =
 \vec{g_{\mathrm{grav}}} + \vec{g_{\mathrm{rot}}} + \vec{g_{\mathrm{rad}}}$.
  This equation has two roots \citep{MMVI}. The first that is met determines the critical 
velocity. The first root is as usual  $v_{\mathrm{crit, 1}} = 
\left( \frac{2}{3} \frac{GM}{R_{\mathrm{pb}}} \right)^{\frac{1}{2}}$, where
$R_{\mathrm{pb}}$ is the polar radius at break--up.
 The second root  $v_{\mathrm{crit, 2}}$ applies to Eddington factors bigger than 0.639. It is
equal to 0.85, 0.69, 0.48, 0.35, 0.22, 0 times $v_{\mathrm{crit, 1}}$ for $\Gamma=$
0.70, 0.80, 0.90, 0.95, 0.98 and 1.00 respectively.

\begin{figure}
\plottwo{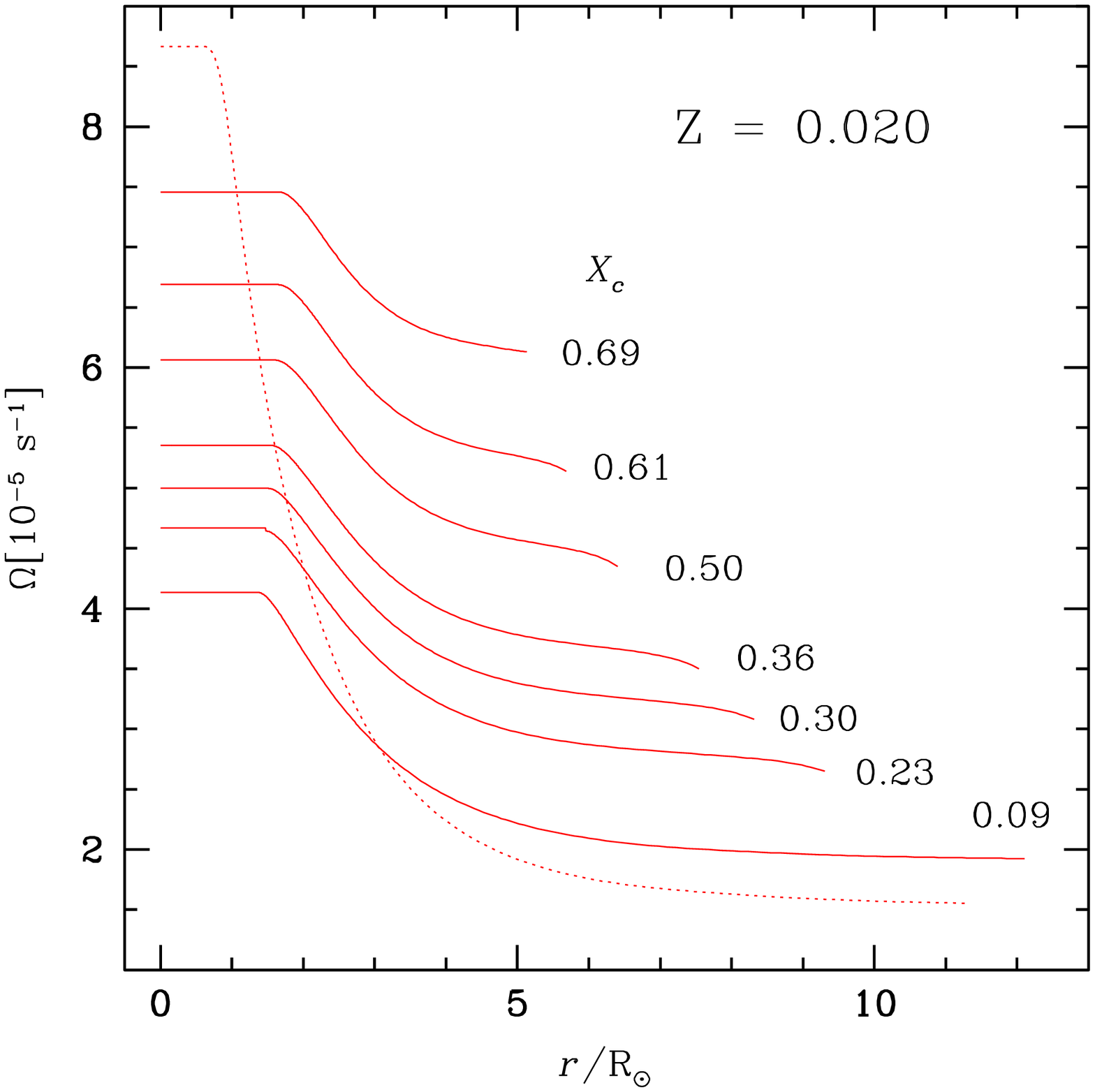}{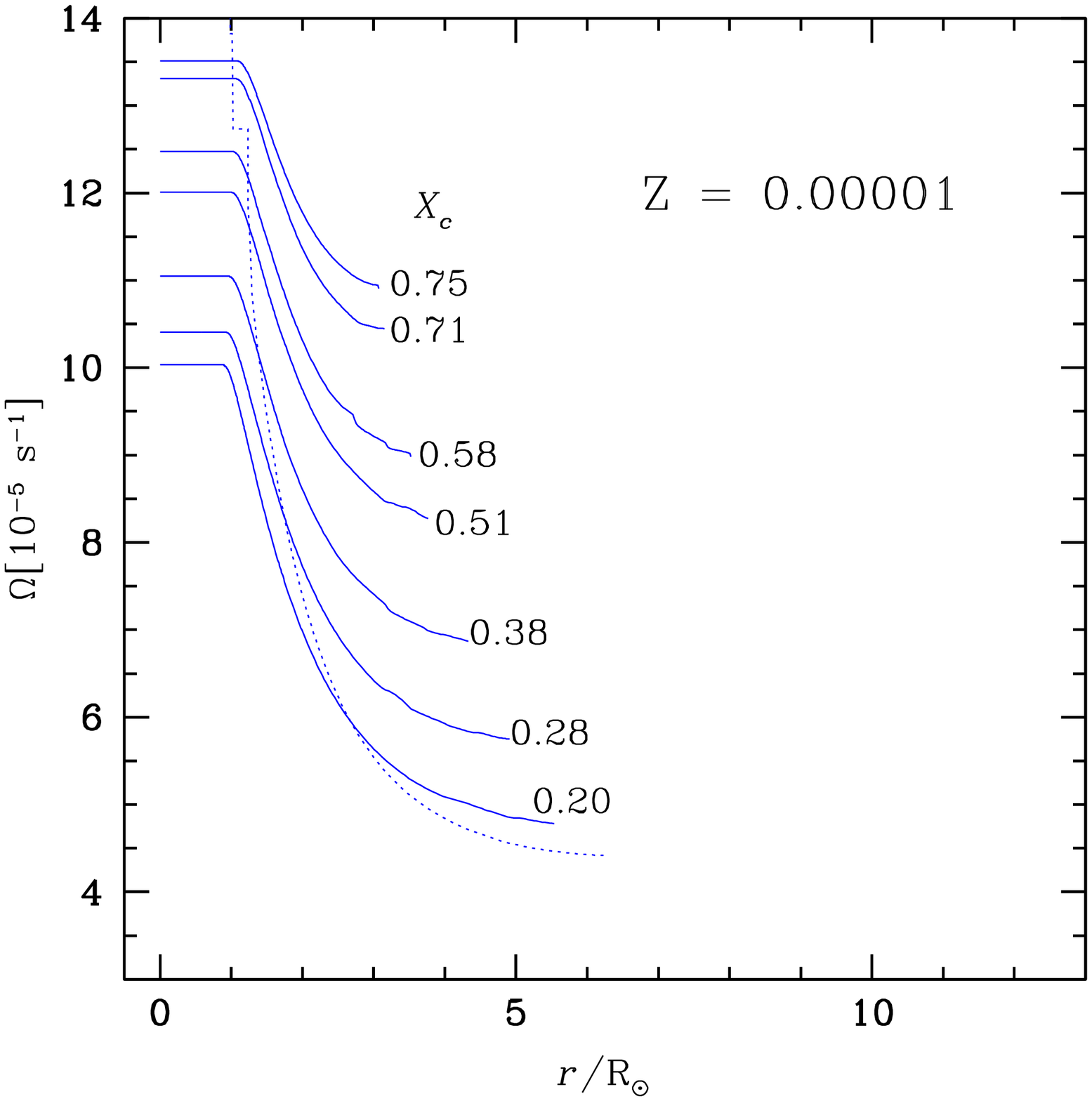}
\caption{Left: Evolution during the MS phase
of the angular velocity $\Omega$  in the interior of a 20 $M_{\odot}$
star at $Z$=0.02 \citep{MMV}. Right: the same for a 20 $M_{\odot}$ star at  $Z=10^{-5}$. The values of
the central H--content $X_{\mathrm{c}}$ are indicated
\citep{MMVIII}.}
\end{figure}

The theory of radiative winds applied to a  rotating star 
leads to an expression  of the mass flux as a function of 
colatitude. We may distinguish two effects. a) The ``$g_{\mathrm{eff}}$--effect'': 
the higher effective gravity makes a higher $T_{\mathrm{eff}}$ at the pole and  favours polar ejection.
For a star hot enough to have electron scattering opacity as the dominant opacity
source from pole to equator, the iso--mass loss curve has a peanut--like 
shape. b) The ``opacity--effect'': 
if the  $T_{\mathrm{eff}}$ of the star is lower than about 25 000 K,  a bistability limit
i.e. a steep increase of the opacity \citep{Lamers95} may occur somewhere between the 
pole and the equator, due to the decrase of $T_{\mathrm{eff}}$ from pole
to equator. This ``opacity--effect'' produces an equatorial enhancement
of the mass loss. The anisotropies of mass loss influence the loss of angular
momentum, in particular polar mass loss removes mass but relatively little 
angular momentum. This may strongly influence the further evolution. 
We may estimate the mass loss rates of a rotating star 
compared to that of a non--rotating star at the same location in the HR
diagram. The result is \citep{MMVI}
\begin{equation}
\frac{\dot{M} (\Omega)} {\dot{M} (0)} \simeq
\frac{\left( 1  -\Gamma\right)
^{\frac{1}{\alpha} - 1}}
{\left[ 1 - 
\frac{4}{9} (\frac{v}{v_{\mathrm{crit, 1}}})^2-\Gamma \right]
^{\frac{1}{\alpha} - 1}} \; ,
\end{equation}
\noindent
where $\Gamma$ is the electron scattering opacity for a non--rotating
star with the same mass and luminosity, $\alpha$ is a force multiplier \citep{Lamers95}. 
For B--stars far from the Eddington Limit, $\frac{\dot{M} (\Omega)} {\dot{M} (0)} \simeq 1.5$
For stars close to
$\Gamma=1$ the  increase of the mass loss rates may reach orders of magnitude. 
\begin{figure}[!ht]
\plotfiddle{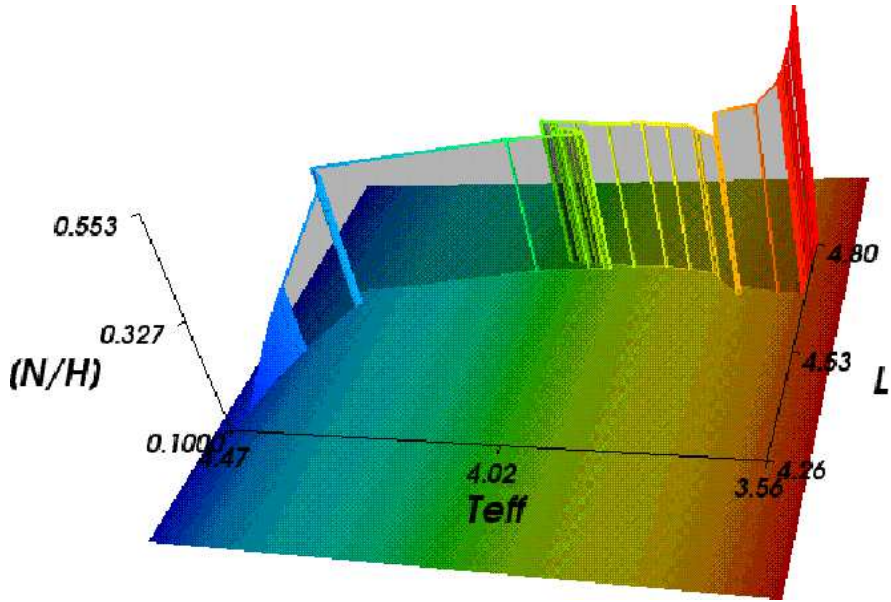}{8.8cm}{0}{50}{50}{-200}{-70}
 \caption{3--D diagram showing the HR diagram  horizontally
  and $N/H$ on the vertical axis. The values of $N/H$ in number 
   are to be multiplied by 10, they are normalised to the value on the ZAMS. Case of a
  15 $M_{\odot}$ model  with $v_{\mathrm{ini}}= 300 $ km/s.}
\end{figure}

On the whole, there are 3  cases of stellar break--up: 1.-- The
$\Gamma$--Limit, when radiation effects largely dominate; 2.-- The $\Omega$--Limit, when
rotation effects are essentially determining break--up; 3.-- The $\Omega \Gamma$--Limit, when both
rotation and radiation are important for the critical velocity. 

\begin{figure}
\plottwo{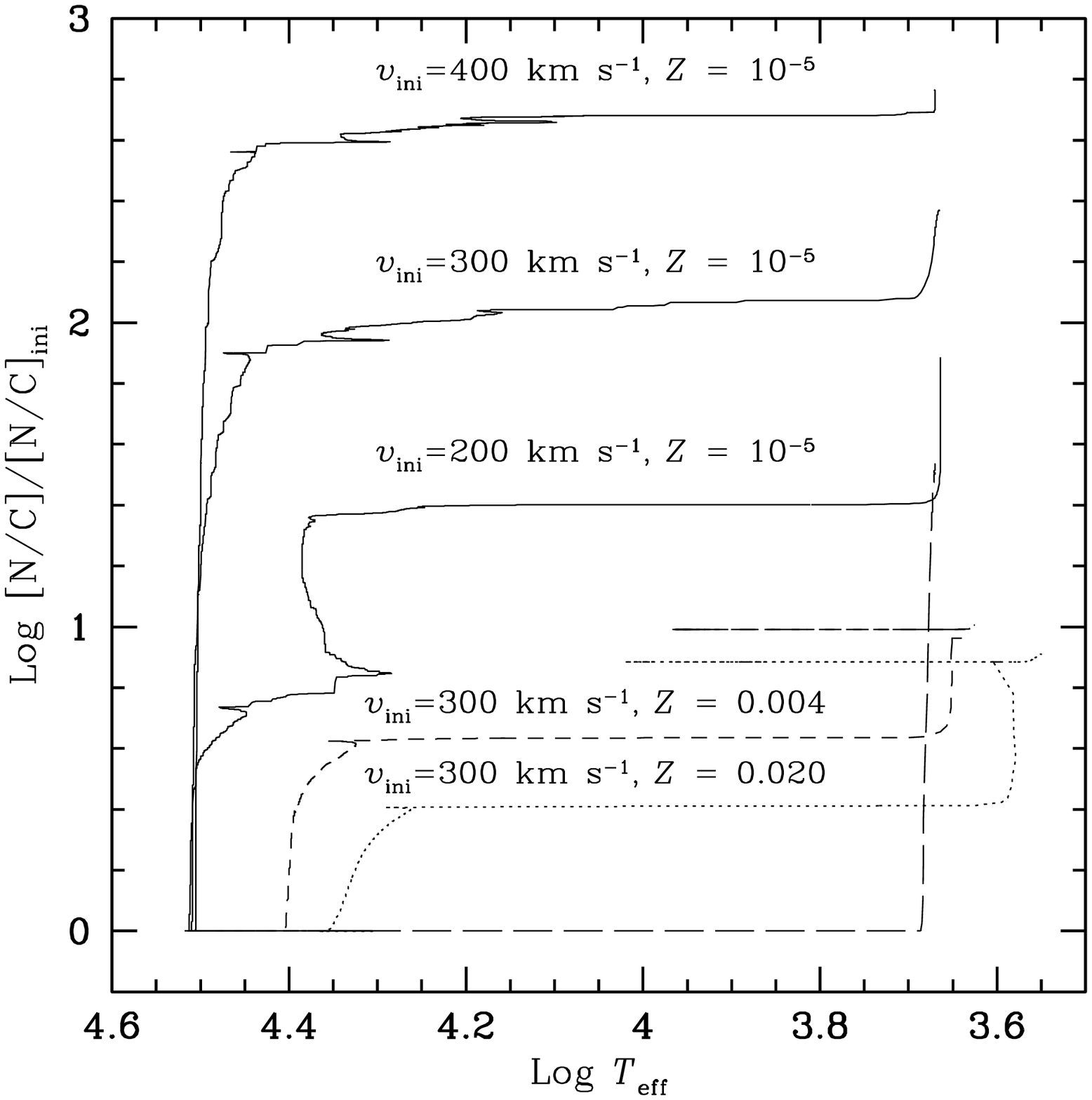}{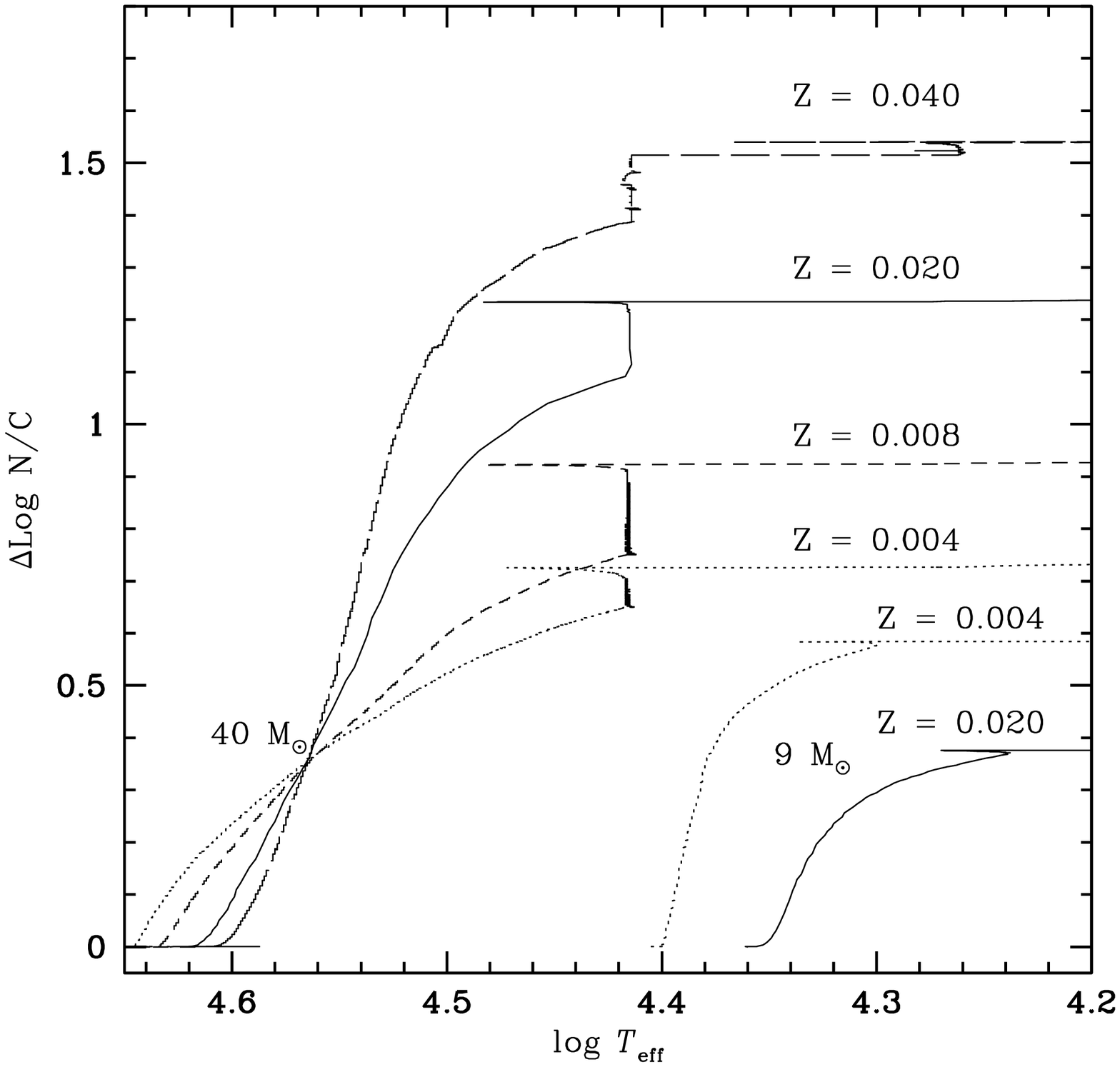}
\caption{Left: Evolution of the ratio $N/C$ (in number) of nitrogen to carbon for 
9 $M_{\odot}$ models of different $Z$ and velocities.
 The ratio are normalized to their initial values. The long--dashed line at the
 bottom corresponds to a non-rotating model of 9 $M_{\odot}$ at $Z=10^{-5}$.
 Right:  the same at the surface of 
 9 and 40 $M_{\odot}$ models of different $Z$.}
\end{figure}

\section{General results}

Grids of models  have been made \citep{MMXI} at $Z$ = 0.020, 0.004 and $10^{-5}$.
Fig.~1 illustrates the HR diagram for non--rotating and rotating stars at solar metallicity.
The rotating models have an initial velocity $v_{\mathrm{ini}}$ of 300 km/s,
which gives an average velocity of 220 km/s during the MS phase, corresponding
to observations. We notice the following points:
\begin{itemize}
\item Rotation increases  the MS lifetime with respect to non--rotating models (up to
 about 40 \%).
\item The values assigned from isochrones with an average rotation velocity  typically lead to ages 25\% larger than without rotation.
\item Rotation strongly affects the lifetimes as blue and red supergiants (RSG). 
In particular in the  SMC, the high observed number of RSG can only
be accounted for by rotating models \citep{MMVII}.
\item Steeper gradients of internal rotation $\Omega$ are built at lower $Z$.
Fig.~2  shows the evolution during the MS phase of $\Omega$
in  models at $Z=0.02$ (left)
and at $Z=10^{-5}$ (right). The steeper $\Omega$--gradient at lower $Z$  favours mixing.
There are 2 reasons for the steeper $\Omega$--gradients. 
One is the higher compactness of the 
star at lower $Z$. The second one is more subtle.  At lower $Z$, the density of the outer layers
is higher, thus the Gratton--\"{O}pik term
is less important. This produces less outward transport of angular momentum and 
contributes to steepen 
the $\Omega$--gradient.
\item  At lower $Z$, rotating stars more easily reach break--up velocities and may
stay at break-up for a substantial fraction of the MS phase.
\item There are various filiation sequences for massive stars, as shown below. 
\end{itemize}

\noindent
{\bf \underline{{$M >90  M_{\odot}$}}}:  O –- Of –- WNL –- (WNE) -– WCL –- WCE -– 
SN (Hypernova low Z  ?)\\
\noindent
{\bf \underline{{$60-90 \; M_{\odot}$}}}: O –- Of/WNL$\Leftrightarrow$LBV -– WNL(H poor)-– WCL-E -– SN(SNIIn?)\\
\noindent
{\bf \underline{{$40-60 \; M_{\odot}$}}}: O –- BSG –-  LBV $\Leftrightarrow$ WNL -–(WNE) -- WCL-E –- SN(SNIb) \\
\hspace*{5.9cm}  - WCL-E - WO – SN (SNIc) \\
\noindent
{\bf \underline{{$30-40 \; M_{\odot}$}}}:  O –- BSG –- RSG  --  WNE –- WCE -– SN(SNIb)\\
\hspace*{4.0cm}                        OH/IR $\Leftrightarrow$ LBV ? \\
\noindent
{\bf \underline{{$25-30 \; M_{\odot}$}}}: O -–(BSG)–-  RSG  -- BSG (blue loop) -- RSG  -- SN(SNIIb, SNIIL)\\
\noindent
{\bf \underline{{$10-25 \; M_{\odot}$}}}: O –-  RSG -– (Cepheid loop, $M < 15 \; M_{\odot}$) – RSG -- 
SN (SNIIL, SNIIp)\\ 
 
\begin{figure}[!ht]
\plottwo{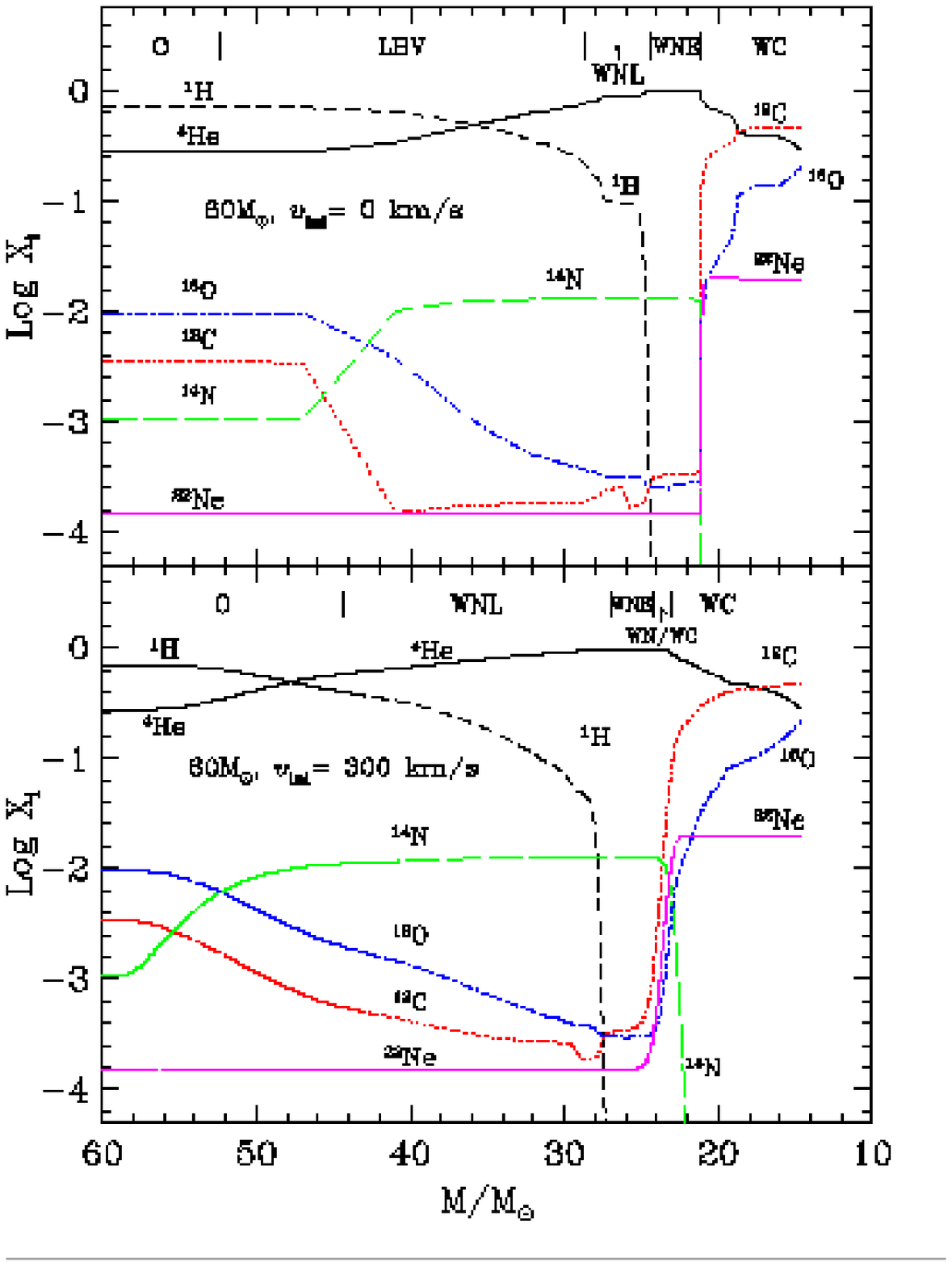}{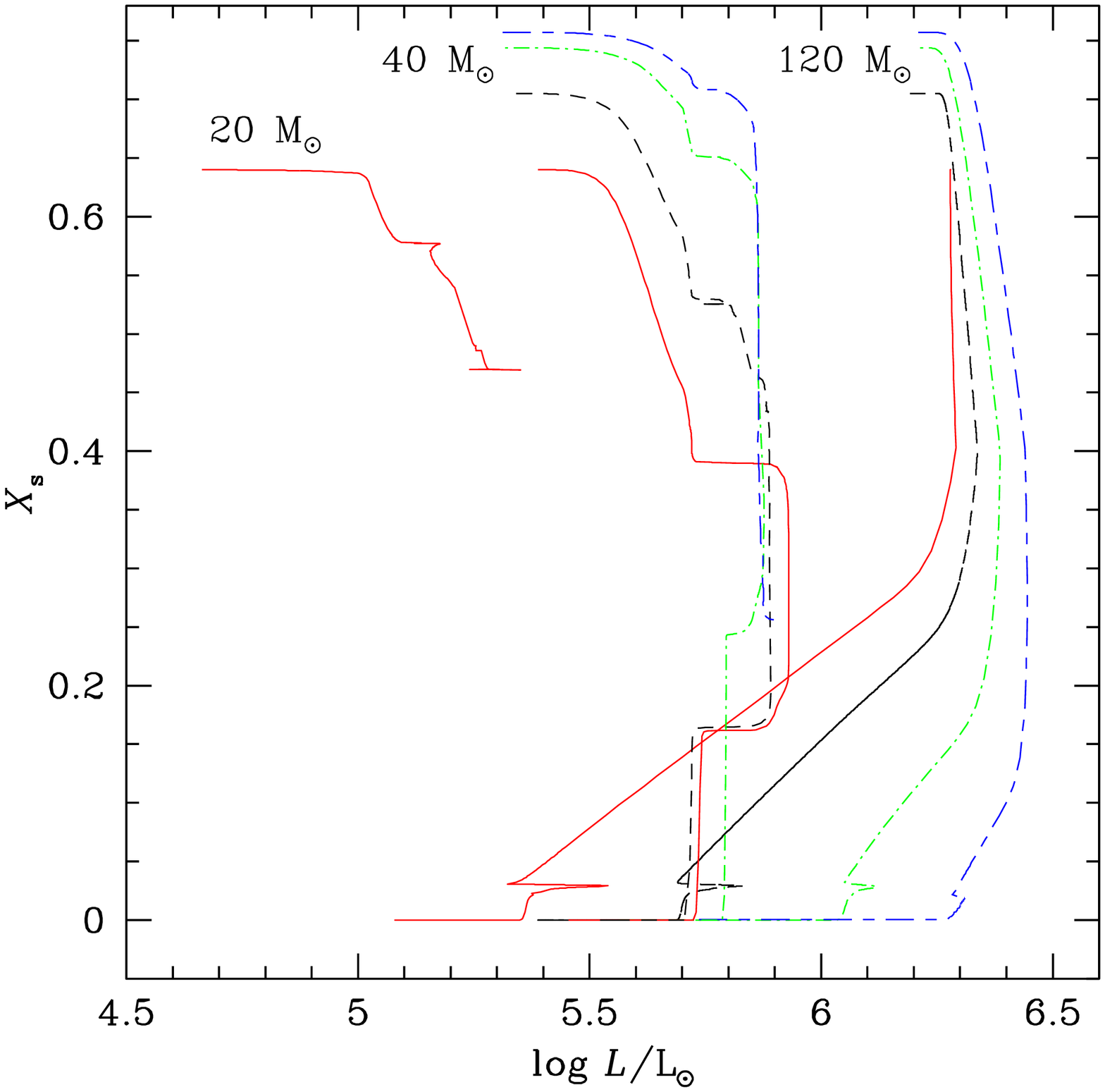}
\caption{Left: Evolution as a function of the actual mass of the 
abundances (in mass fraction) at the surface of a non--rotating  
(upper panel) and a rotating (lower panel) 60 M$_\odot$ stellar model.
Right: Evolutionary tracks in the $X_{\rm s}$ versus log {\it L}/ {\it L}$_\odot$ plane,
where $X_{\rm s}$ is the hydrogen mass fraction at the surface. The initial masses are indicated.
Long--short dashed curves show the evolution of $Z$ = 0.004 models, 
dashed--dotted curves, short--dashed curves
and continuous lines show the evolutions for $Z$ = 0.008, 0.020 and 0.040 respectively. }
\end{figure}
\noindent
The sign $\Leftrightarrow$ means back and forth motions between the two  stages. The limits 
between the various scenarios  depend on metallicity $Z$ and rotation.
  The various types of supernovae are tentatively indicated.

  \section{Evolution of  surface abundances in massive stars}
  \subsection{N/C enrichments during the MS phase at various metallicities}

\begin{figure}[!ht]
\plotfiddle{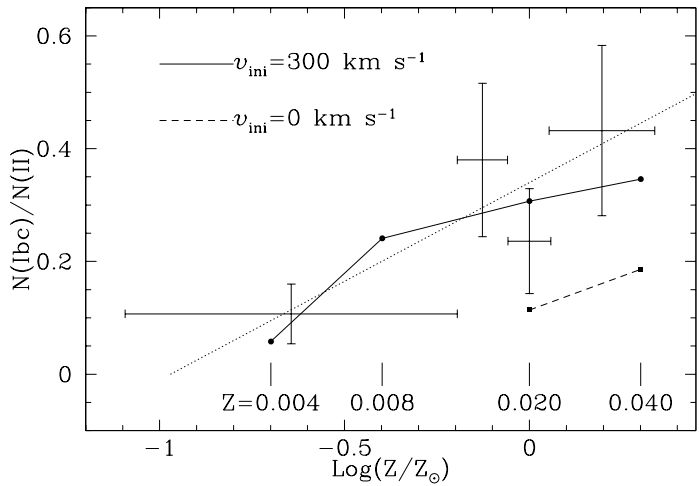}{6.2cm}{-90}{34}{34}{-150}{200}
\caption{Variations of the number ratios of SN types Ib and Ic with respect to SNII. The crosses
with error bars correspond to observational data \citep{PrantzosB03}. The dotted line is a fit proposed by these authors. The continuous and dashed lines show the predictions of rotating and non--rotating
models.}
\end{figure}

As a result of the rotational mixing, products of the CNO processing are reaching the stellar surface 
during MS evolution and produce $N/C$ enhancements, as 
observed \citep{GiesL92, Herrero92, Lyub96} since long in OB--stars. Fig.~3 illustrates 
in a 3--D plot the growth of N/H  on the track   of a  15 $M_{\odot}$
rotating with an initial rotation $v_{\mathrm{ini}}=300$ km/s. This velocity corresponds to the
average observed velocity during the MS phase of OB stars of about 220 km/s. We see that the 
N--enrichment progressively occurs during the MS phase, then it keeps about constant 
during the crossing of the HR and rises up again due to the convective dredge--up in red supergiants.

  Fig.~4 left shows the evolution of the $N/C$ ratios  in models of
rotating stars with 9 $M_{\odot}$ for initial $Z$ = 0.02, 0.004 and  10$^{-5}$.
At zero rotation, there is no enrichment during 
the MS phase (except at $Z$=0.02 for M $\geq$ 60 $M_{\odot}$ due to very high
mass loss). At 9 M$_{\odot}$ for  solar $Z$ and  an initial rotation of 300 km s$^{-1}$, 
the $N/C$ ratio  increases by about 0.4 dex  during the
MS phase. The \emph {relative} values of the $N/C$ ratios
increase with decreasing $Z$, in particular $N/C$ increase by two orders of a magnitude
for   $Z$ = 10$^{-5}$. This results from the steeper $\Omega$--gradients and greater compactness. 
 Of course, large $N/C$ enhancements are  accompanied by  small enrichments in helium, 
typically of a few hundredths. The larger $N/C$ enrichments at lower $Z$ have been nicely
confirmed by abundances determinations in the SMC \citep{VennP03}. They found relative $N/H$ enrichments 
for A--type supergiants in the SMC reaching an order of magnitude or more,
 while in the Milky Way the enrichments are only by a factor 2 to 3. 

 During the He--burning phase, the rotation velocities become all the same whatever the initial
rotational velocities. Thus, in the He--burning
phase we  may have very different surface chemical compositions
for actually similar rotation velocities. 

Fig.~4 right compares the relative $N/C$ enrichments for initial 40 and 9 $M_{\odot}$ stars.
For high mass stars ($\geq 30$ $M_{\odot}$), the higher $Z$ models have the larger enrichments, i.e.
a situation  opposite to that 
described above; the reason is that  mass loss effects dominate over mixing. Higher $Z$ stars have a higher mass loss, which peels off the stars and make the products of  CNO process visible
at the stellar surface.

\subsection{The abundances in Wolf--Rayet stars}

\begin{figure}[!ht]
\plotfiddle{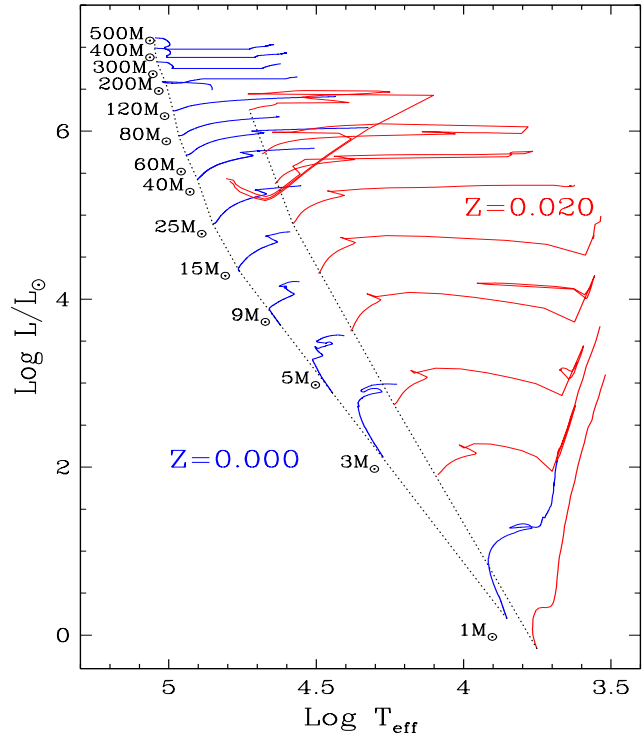}{9.2cm}{0}{45}{37}{-150}{-20}
\caption{Evolutionary tracks for models at solar and $Z=0$ \citep{Feijoo99}. }
\end{figure}
For WR stars, the first major constraint is to explain their relative number ratios 
 to O--type stars, 
which  strongly decrease for lower  $Z$. Rotation
 makes massive stars to enter earlier in the WR phase and this also  increases
the lifetimes as WR stars. When account is given to both rotation
and to the dependence of the mass loss rates to metallicity $Z$, the comparisons with
 observations are greatly improved  \citep{MMXI}.


Fig.~5 left shows the evolution of surface abundances for a 60 $M_{\odot}$ star 
with and without rotation. Rotation makes  smoother changes of abundances, due to internal mixing.
For WN stars, the transition  phase, when they still have H present,
 becomes  longer due to rotation and this increases
the late WN phase (WNL), where H is usually present. The CNO abundances  at the end of the
WNL phase are the same for rotating and non--rotating models, because  they 
are model independent and  determined just by CNO nuclear equilibrium.
The same is true for the early WN (WNE), which in general have no or little H present.
Indeed, CNO abundances in WN stars provide a unique test of the physics of the CNO cycle.

For the WC phase, the milder composition gradients, when revealed at the surface,
 make smoother transitions with lower C/He ratios, in good agreement 
with observations \citep{Crowther95}. The abundances in WC stars are not equilibrium values, but are products of the partial He--burning, thus they are model dependent and offer 
a most interesting  test on the stellar models.
Thus, the abundances in WN and WC stars tell us different stories.

\begin{figure}[!ht]
\plotfiddle{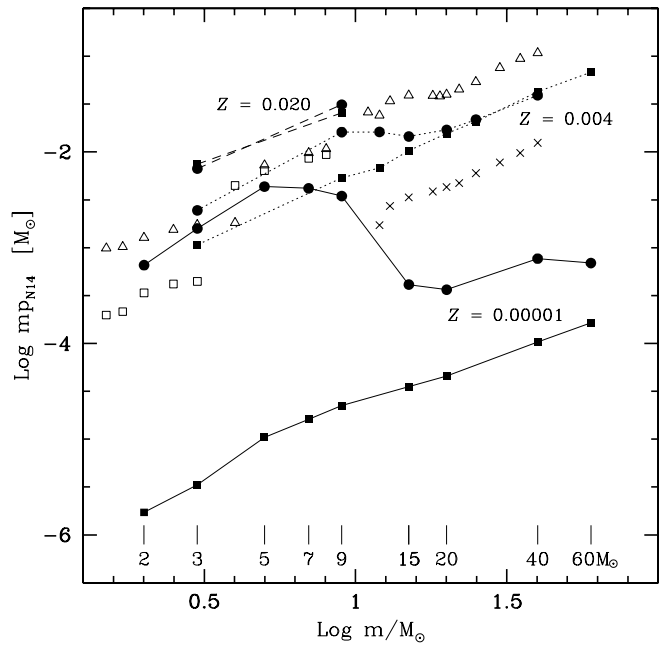}{8.0cm}{0}{45}{45}{-150}{-80}
\caption{Variation as a function of the initial mass of the stellar yields in $^{14}$N for different
metallicities and rotational velocities. The continuous lines refer to the models 
at $Z =10^{-5}$, the dotted lines show the yields from  models 
at $Z$ = 0.004  \citep{MMVII}, the dashed lines present the yields for solar $Z$
models. The filled squares and circles indicate the cases
without and with rotation respectively ($v_{\rm ini}$ = 300 km s$^{-1}$). 
The crosses are   models   at $Z=0.1 \; Z_{\odot}$  \citep{WW95},    the 
empty squares  at $Z=0.004$ \citep{ho97} and the empty triangles are for solar $Z$ models
\citep{ho97} up to 8 $M_{\odot}$  with complements \citep{WW95}.
 }
\end{figure}

Fig.~5 right shows the evolution of the H--surface content $X_{\mathrm{s}}$ vs. luminosity.
This is a very constraining diagram especially for the transition stages from O, Of, LBV to
WN stages. It is useful for establishing the proper filiations between  such stars. It clearly 
supports the view that there may be back and forth evolution between LBV and WNL stars, and that in general
WN stars succeeds the LBV stage.
We also see that descendants from high masses at higher $Z$ significantly decrease
in luminosity during their  evolution.

The rare WO stars, characterized by a high $O/C$ ratio represent a more advanced stage of nuclear
processing. Curiously enough, such stars which may be the progenitors of supernovae SNIb/c
are found  only at lower $Z$. The physical reasons of that have been explained \citep{SmithM91}:
lower $Z$ stars become WC stars (if they do it) only very late in evolution, 
i.e.~with a high $O/C$ ratio. 
At the opposite, at higher $Z$ the WC stars may occur at an early stage of He--processing, i.e.
with a low $O/C$ ratio. Fig.~6 comapres the observed and predicted variations 
of the number ratio of SNIb or c and SNII and shows an interesting agreement.
This is noticeable  because of the possible
connection WO stars - SNIb/c - GRBs ($\gamma$ Ray Bursts). In this context,
we consider WO stars as good candidate for GRB progenitors.

\begin{figure}[!ht]
\plotfiddle{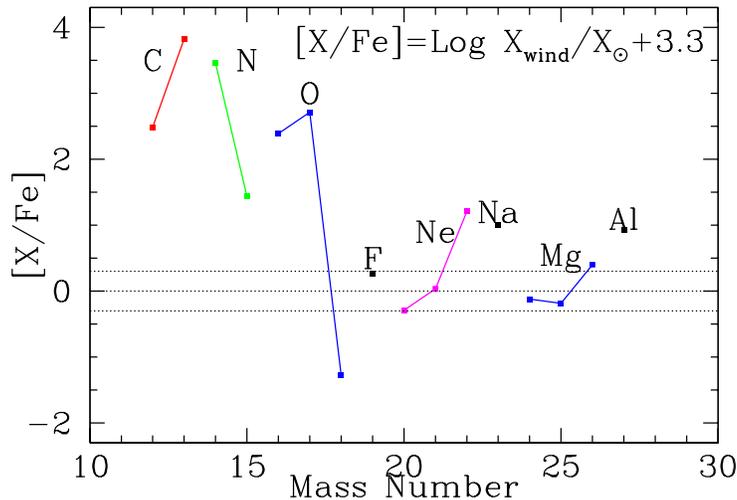}{6.5cm}{-90}{35}{35}{-175}{210}
\caption{Chemical composition of the matter ejected through stellar winds by 
a 60 M$_\odot$ at $Z$ = 0.00001 with an initial velocity equal to $2/3$ of the
break--up velocity. Note that
 log ($Z_\odot/Z$)=3.3 in this diagram. The horizontal dotted lines show the
solar ratios (middle) and a factor two enhancement/depletion.}
\end{figure}
\section{Abundances and yields in very low $Z$ and $Z=0$ stars}
Evolutionary models at $Z=0$ behave very differently from models at solar composition, 
as shown by Fig.~7, in particular the evolutionary scenarios and final stages  are different
\citep{hegerW02}. Let us concentrate on the chemical abundances. For metallicities 
lower than that of the SMC, the internal mixing in intermediate and massive stars 
is sufficient to bring new C from the He--burning core to the H--shell where it is then
turned to N. This nitrogen is called ``primary'' as it does not result
 from CNO elements initially present. Fig.~8 shows the yields in N as a function of the initial mass for $Z=0.02,\; 0.004$ and $10^{-5}$. At $Z=0.02$ and $Z=0.004$ the level of N production
 is just that resulting from the initial CNO elements, even if at $Z=0.004$ the  N--enrichment is
 somehow larger than at $Z=0.02$ (cf. Sect. 4.1). However, at $Z=10^{-5}$ there is an
  overproduction of N 
  by 1 or 2  orders of magnitude. This primary N is mainly produced in intermediate
  mass stars, but the contribution of massive stars is also significant. 

Remarkably, at very low $Z$ the fast rotating stars
of intermediate masses  which reach the TP--AGB phase
(this occurs for M $\leq 7 \; {\rm M}_{\odot}$) 
get a higher $Z$ during this phase due to their enrichment in 
CNO elements. As an example, a  7 M$_{\odot}$ has 
 X(CNO) = 3.1 $\cdot 10^{-3}$ which is 430 times
the initial CNO content. Thus very low $Z$ stars
may get a higher $Z$ value in post MS phase, which may drive  large mass loss rates.

\begin{figure}[!ht]
\plotfiddle{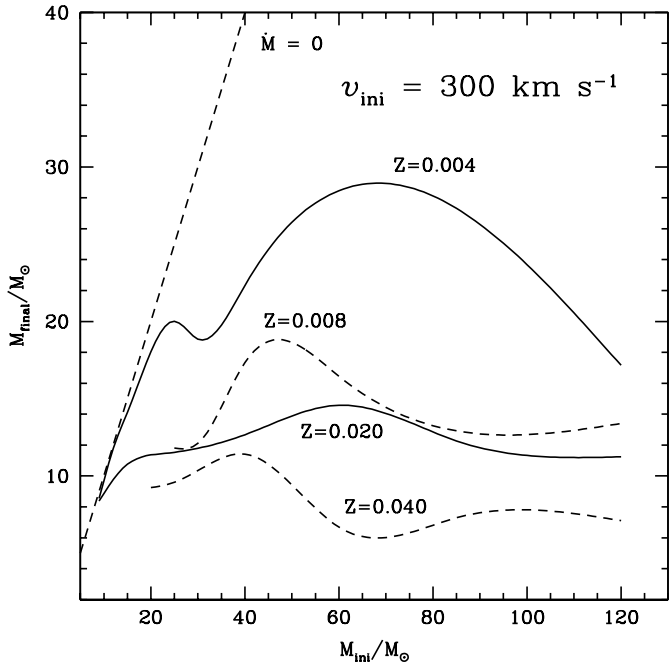}{8.5cm}{0}{46}{46}{-150}{-75}
\caption{Relation between the final and the initial mass for rotating models at various $Z$.
The final masses of models at very low and zero $Z$ are still uncertain, but
they probably significantly differ from the line of those with zero mass loss
rate $\dot{M}=0$. }
\end{figure}

Intermediate and massive stars
of low $Z$ easily reach break--up velocities during a significant fraction of their MS phase.
This results from the growth of the rotation velocity at their surface, due to the very weak
mass loss and to the mild internal
coupling (due to meridional circulation) with the contracting core, which spins faster and faster
during evolution. 
As a result of these 2 effects (mass losses at break--up and from 
higher surface $Z$), massive rotating stars of very low $Z$ also lose a large fraction of
their masses. As an example, a star with an initial mass of 60 $M_{\odot}$
at $Z=10^{-5}$ finishes its life as a 30 $M_{\odot}$ core and may possibly lead to
 an hypernova forming a black hole.  
The composition of the wind ejecta of this 60 $M_{\odot}$ model is illustrated
 in Fig.~9. It shows large overabundances of C, N and O, as well as lower enrichments
 in Ne, Na and Al from secondary cycles associated to the  CNO cycles.
 Remarkably, these abundance anomalies well correspond to those observed
 at the surface of the  very metal poor halo stars showing C excess.
 If there is some other ejecta at the time of the supernova explosion, this 
 material with anomalous abundances will be diluted in the current ejecta
 with $\alpha$--rich nuclei.
 
 Thus, we see that if very low $Z$ stars rotate at a significant rate, they may
 interestingly contribute to account for some abundance anomalies observed
 in very metal poor halo stars. In this respect, it is interesting to mention that 
 the the ratio  of Be--stars to all B stars is  3 to 4 times higher
 in the SMC than in the solar neighbourhood \cite{MGM99}. This suggests that the fraction of fast 
 rotating stars is higher at lower $Z$.
 \section{Chemical yields from rotating stars}
  
 \begin{figure}[!ht]
\plottwo{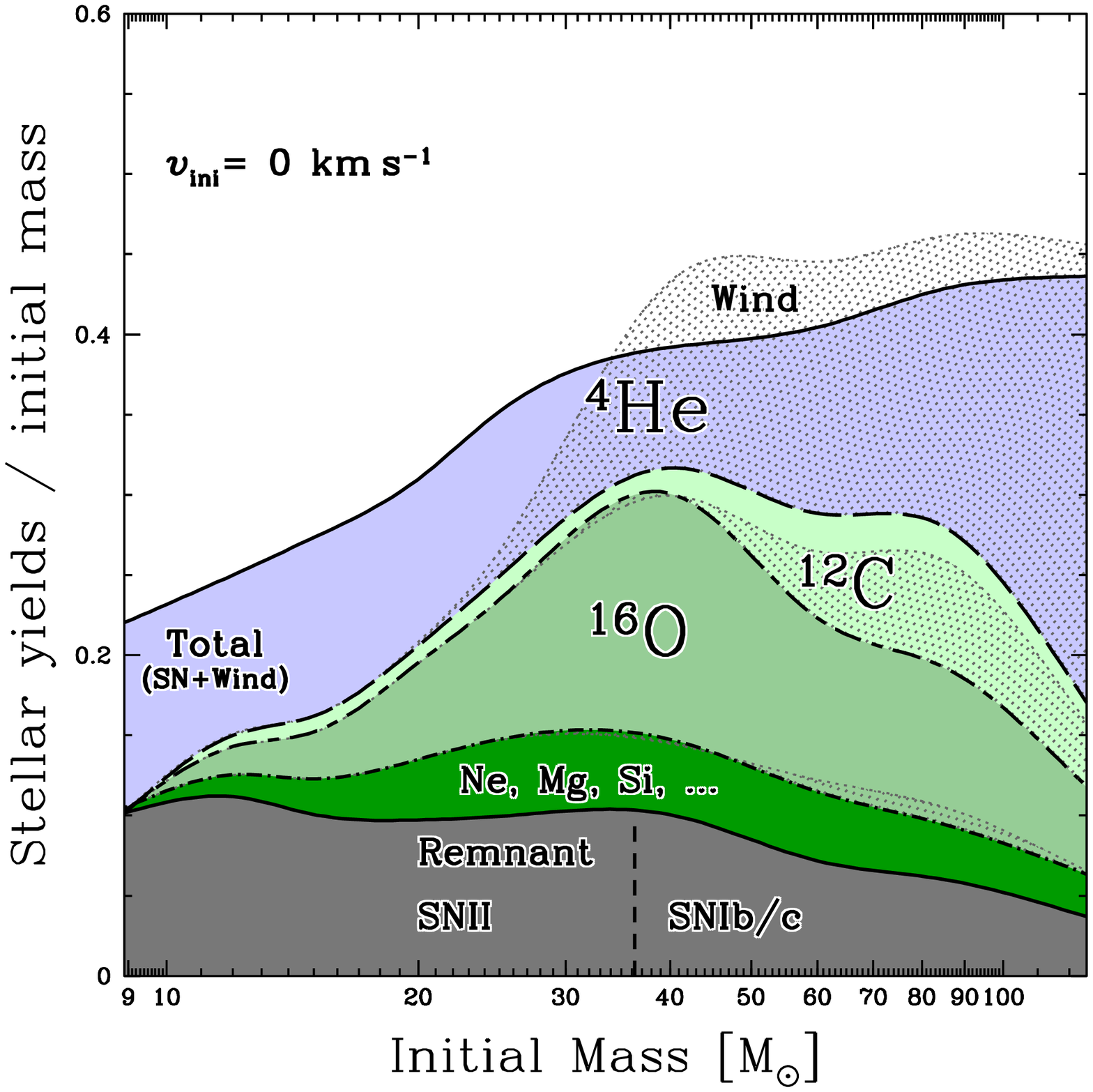}{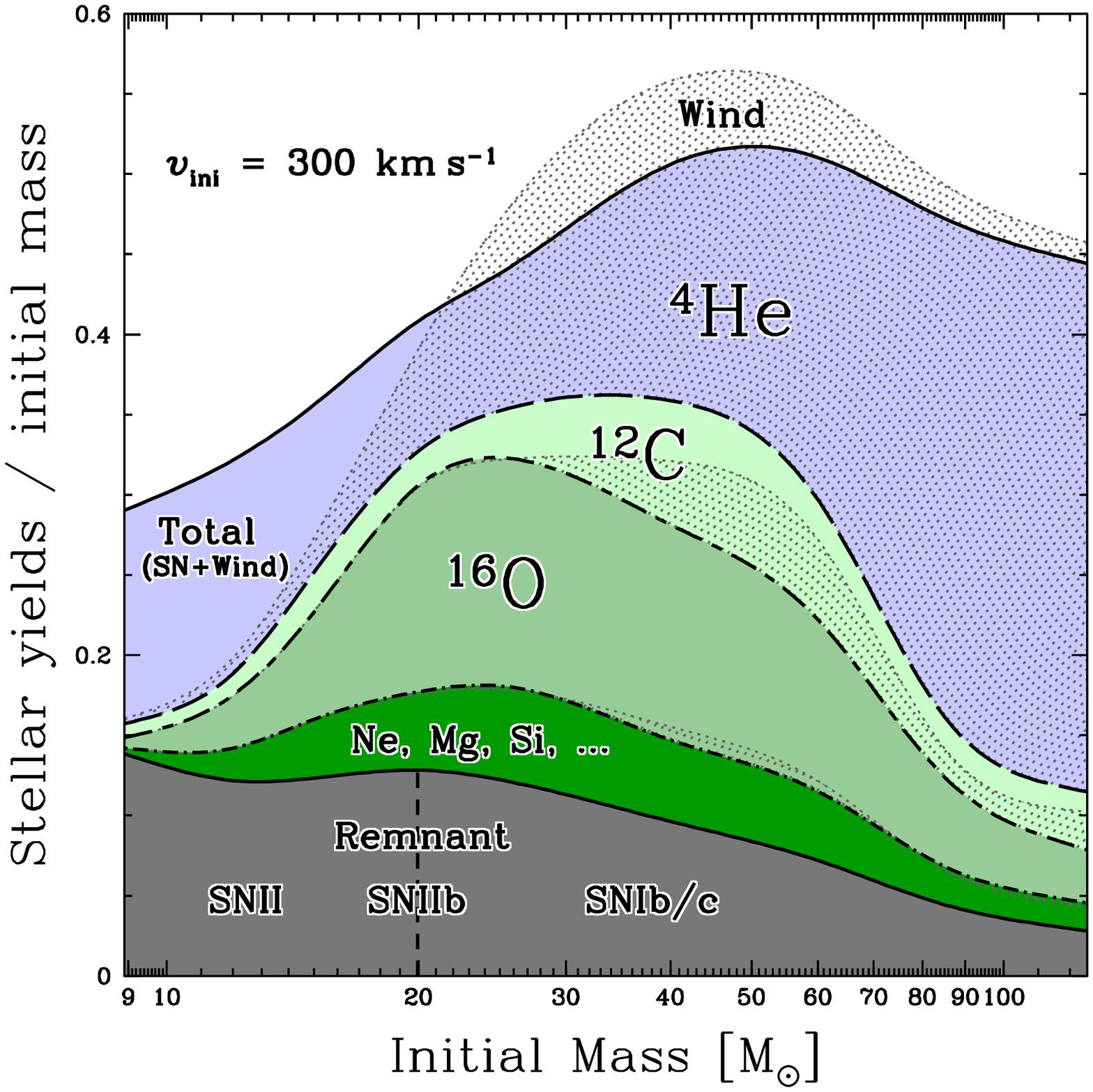}
\caption{Left: The chemical yields  for models without rotation. Right: The chemical yields
for models with $v_{\mathrm{ini}}= 300$ km/s \citep{HMMXII}.}
\end{figure}
\begin{figure}[!ht]
\plottwo{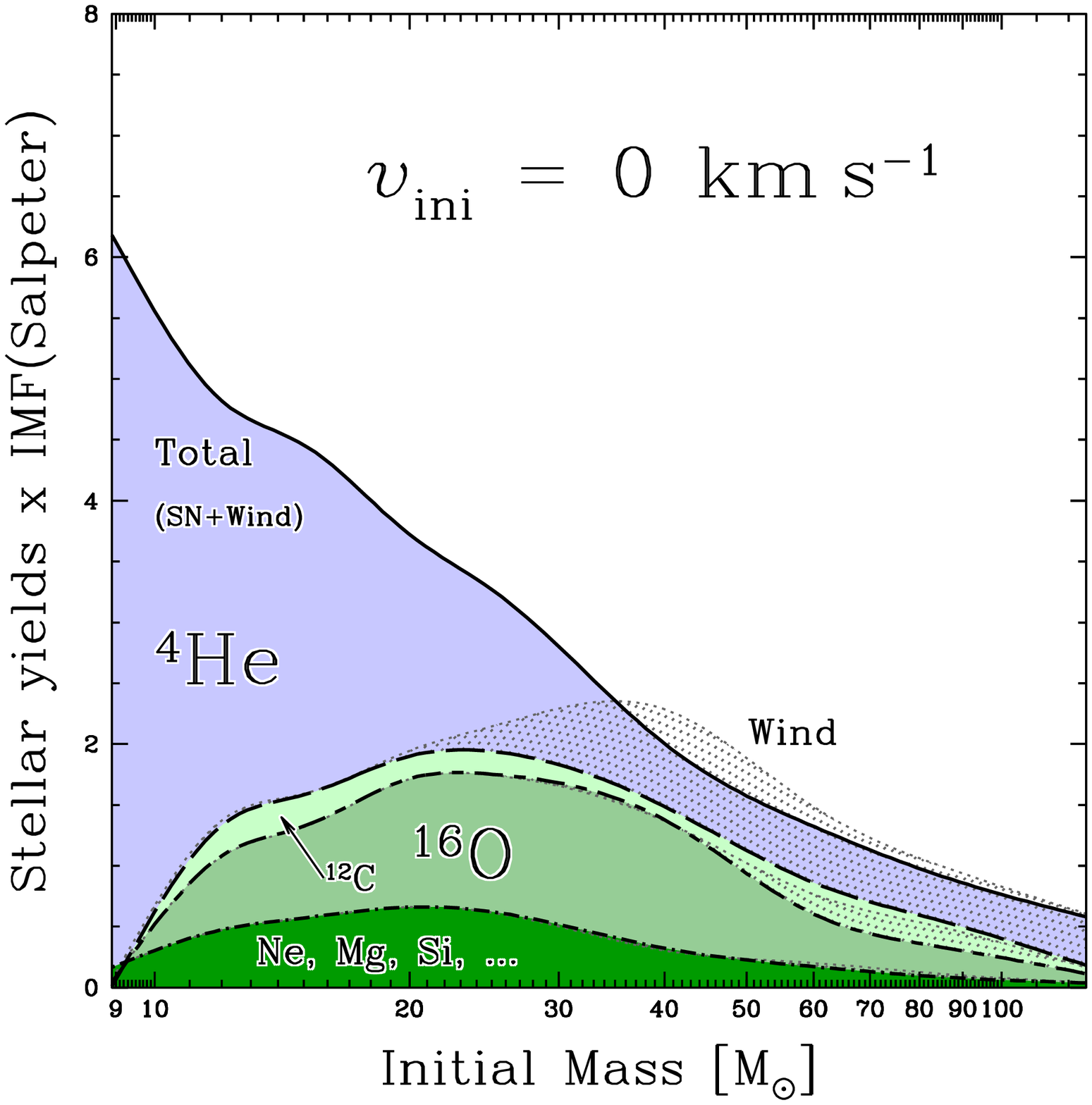}{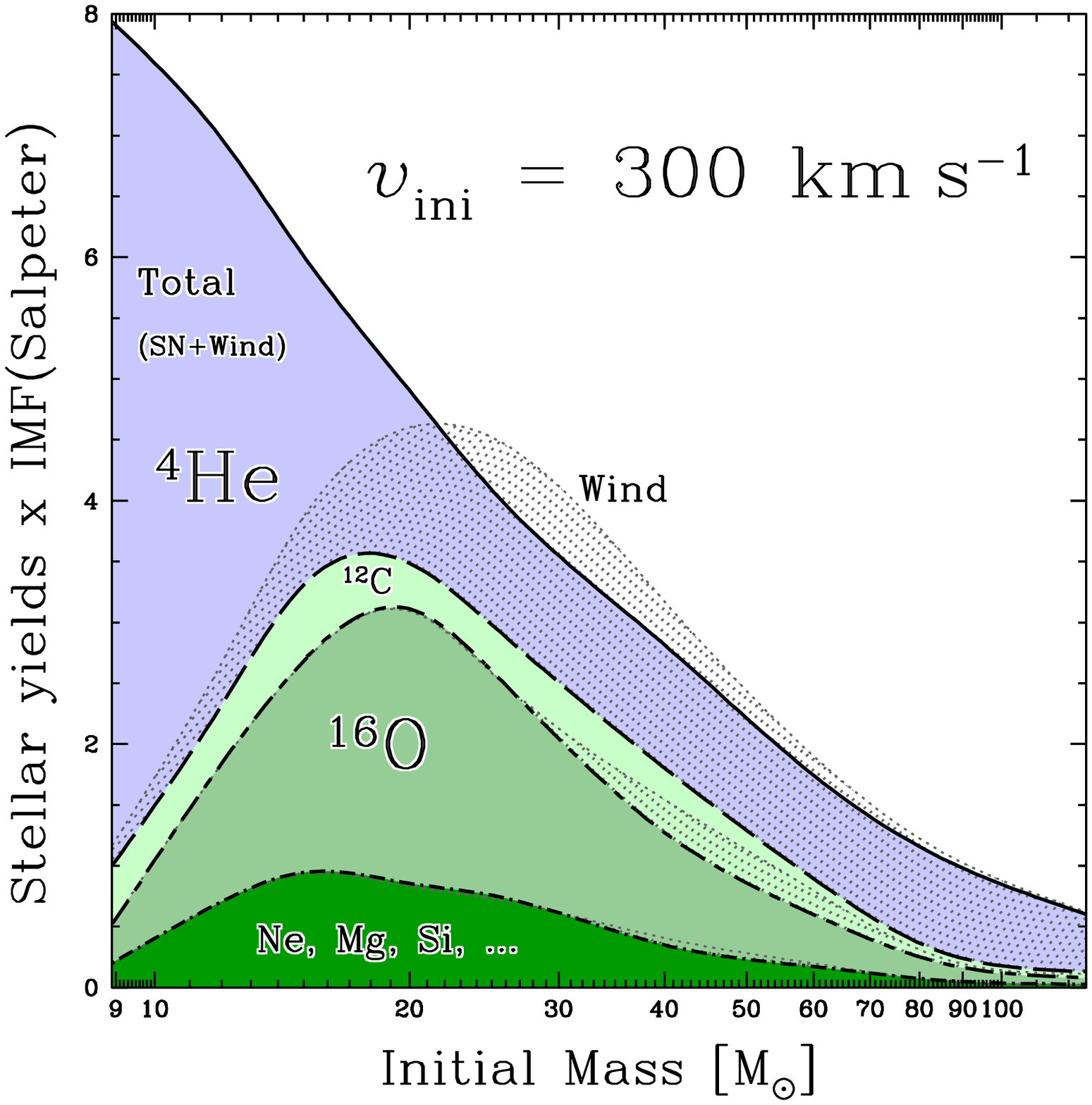}
\caption{Left: The yields x IMF for models without rotation. Right: The yields x IMF
for models with $v_{\mathrm{ini}}= 300$ km/s \citep{HMMXII}.}
\end{figure}
 
 The relations for various metallicities $Z$ between the final masses at the time of 
 supernovae explosion and the initial stellar masses are still uncertain due
 to the uncertainties in mass loss rates.
 The present mass loss estimates lead to the relations
  illustrated in Fig.~10 for  $Z \geq$ 0.004. The models are based 
 on the expression of the mass loss rates as a function of metallicity given
 by $\dot{M}(Z)= \dot{M}(Z_{\odot})\;(Z/Z_{\odot})^{\alpha}$ \citep{Kudr02}, a
  relation which is certainly different in the various evolutionary phases, 
 since the mechanisms of mass loss are not the same.
 For the very low and zero $Z$, the uncertainties are still too large,
  so that we do not know the final masses.
 However, we emphasized above that, in very low $Z$ stars,
 surface enrichments due to internal mixing may enhance the mass loss by stellar winds,
 also  the stars may reach break--up and experience significant mass loss.
 The values of the final masses are critical for determining the types of supernovae
 and also for nucleosynthesis, because as is evident what is escaping in the winds
 is not further nuclearly processed.

The model evolution of rotating stars has been pursued up to the presupernova stage \citep{HMMXII},
since we know that
nucleosynthesis is also influenced by rotation \citep{HLW00}. Fig.~11
show the chemical yields from models without and with rotation. Fig.~12
 shows these  yields multiplied by the initial mass function (IMF). The main
 conclusion is that below an initial mass of 30 $M_{\odot}$, the cores are larger 
 and thus the production of   $\alpha$--elements is enhanced. Above  30 $M_{\odot}$,
 mass loss is  the dominant effect and more He is ejected before being 
 further processed, while the size of the core is only slightly reduced.
 When we account for  the weighting by the  IMF, the production of oxygen and
 of $\alpha$--elements  is globally enhanced as illustrated by Fig.~12, while the effect on the He--production in massive stars remains limited. It will be interesting to explore the consequences of these new yields for the chemical evolution of the Galaxy.

 As a general conclusion, we see that the chemical abundances at all stages are a most 
 constraining test of stellar evolution and in particular of the internal mixing process
 and of mass loss, which are effects strongly influencing all the outputs of
  evolution and nucleosynthesis.

\end{document}